\documentclass[11pt]{article}
\usepackage{moriond,epsfig}

\def\be{\begin{equation}}
\def\ee{\end{equation}}
\def\bea{\begin{eqnarray}}
\def\eea{\end{eqnarray}}

\begin{document}
\vspace*{4cm}
\title{DIPHOTON SPECTRUM IN THE MASS RANGE 120-140 GEV AT THE LHC}

\author{ L.J CIERI }

\address{INFN, Sezione di Firenze, Via Sansone 1, I-50019 Sesto Fiorentino, Florence, Italy
}

\maketitle\abstracts{
We consider direct diphoton
production in hadron collisions. We compute the next-to-next-to-leading
order (NNLO) QCD radiative corrections at the fully-differential level.
Our calculation uses the $q_T$ subtraction formalism and it is implemented
in a parton level Monte Carlo program, which allows the user to apply arbitrary kinematical cuts on the
final-state photons and the associated jet activity, and to compute the
corresponding distributions in the form of bin histograms. We present selected
numerical results related to Higgs boson searches at the LHC, and we show how
the NNLO corrections to diphoton production are relevant to understand the main background
of the decay channel $H\rightarrow\gamma \gamma$.}

\section{Introduction}
Diphoton production is a relevant process
in hadron collider physics.
It is both a classical signal within the Standard Model (SM)
and an important background for Higgs boson and new-physics
searches.
Recent results from the LHC indicates that the Higgs boson
mass $m_H$ must be low ($114$~GeV$< m_H < 130$~GeV), and
 thus the preferred 
search mode involves Higgs 
boson production via gluon fusion followed by the rare decay into a pair of photons.
%
We are interested in the process $pp \rightarrow \gamma \gamma X$,
which, at the lowest order in perturbative QCD, occurs \textit{via} 
the quark annihilation subprocess $q\bar{q} \rightarrow \gamma \gamma$. The QCD corrections at 
the next-to-leading order (NLO) in the strong coupling $\alpha_{\mathrm{S}}$ involve the  
quark annihilation channel and a new partonic channel, \textit{via} the
subprocess $qg \rightarrow \gamma \gamma q$. These corrections have been
computed and implemented in the fully-differential Monte Carlo codes
\texttt{DIPHOX},\cite{Binoth:1999qq} \texttt{2gammaMC} \cite{Bern:2002jx} and 
\texttt{MCFM}.\cite{Campbell:2011bn} A calculation that includes the effects of
 transverse-momentum resummation is implemented in 
\texttt{RESBOS}.\cite{Balazs:2007hr}
At the next-to-next-to-leading order 
(NNLO), 
the $gg$ channel starts to contribute,
and
the large gluon--gluon luminosity makes this channel sizeable. 
Part of the contribution from this channel,
the so called {\it box contribution}, was computed long ago \cite{Dicus:1987fk} and 
its size turns out to be comparable  
to the lowest-order result.
%
Besides their {\it direct} production from the hard subprocess, photons can also
arise from fragmentation subprocesses of QCD partons. The computation of
fragmentation subprocesses requires (poorly known)
non-perturbative information, in the form of 
parton  
fragmentation functions of the photon.
The complete NLO single- and double-fragmentation contributions are implemented in \texttt{DIPHOX}.\cite{Binoth:1999qq}
The effect of the fragmentation contributions 
is sizeably reduced by the photon isolation criteria that are 
necessarily
applied in hadron collider experiments to suppress the very large irreducible
background (e.g., photons that are faked by jets or produced by hadron decays). 
The standard cone isolation and the `smooth' cone isolation proposed
by Frixione \cite{Frixione:1998jh} are two of these criteria. The standard cone
isolation is easily implemented in experiments, but it only suppresses a fraction of the 
fragmentation contribution.
The smooth cone isolation (formally) eliminates the entire fragmentation 
contribution, but its experimental implementation is still under study.\cite{CERN-note} However, it 
is important to anticipate (work to appear), that in some kinematical regions (e.g for Higgs boson searches), 
the standard cone and the Frixione
 isolation criteria give basically the same theoretical answer.\footnote{The use of the same parameters
in both criteria is understood.}

\section{Diphoton production at NNLO}
We consider the inclusive hard-scattering reaction $h_1+h_2\to \gamma\gamma +X ,$
where the collision of the two hadrons, $h_1$ and $h_2$,
produces the diphoton system 
$F \equiv \gamma\gamma$ with high invariant mass $M_{\gamma \gamma}$.
The evaluation of the
NNLO corrections to the this process requires the knowledge 
of the corresponding partonic scattering amplitudes
with $X=2$~partons (at the tree level, \cite{Barger:1989yd}) $X=1$~parton (up 
to the one-loop level \cite{Bern:1994fz})
and no additional parton (up to the two-loop level \cite{Anastasiou:2002zn})
in the final state.
The implementation of the separate scattering amplitudes in a complete
NNLO (numerical) calculation is severely complicated by 
the presence of infrared (IR) divergences that occur at intermediate stages. 
The $q_T$ subtraction formalism \cite{Catani:2007vq} is a method that handles
and cancels these unphysical IR divergences up to the NNLO.
The formalism applies to generic hadron collision processes that involve
hard-scattering production of a colourless high-mass system $F$.
 Within that framework,\cite{Catani:2007vq} the corresponding cross section is written as:
\begin{equation}
\label{main}
d{\sigma}^{F}_{(N)NLO}={\cal H}^{F}_{(N)NLO}\otimes d{\sigma}^{F}_{LO}
+\left[ d{\sigma}^{F+{\rm jets}}_{(N)LO}-
d{\sigma}^{CT}_{(N)LO}\right]\;\; ,
\end{equation}
where $d{\sigma}^{F+{\rm jets}}_{(N)LO}$ represents the cross section for the
production of the system $F$ plus jets at (N)LO accuracy~\footnote{In the case of
diphoton production, the NLO calculation of 
$d{\sigma}^{\gamma\gamma+{\rm jets}}_{NLO}$ was performed in 
Ref.\cite{DelDuca:2003uz}}, and
$d{\sigma}^{CT}_{(N)LO}$ is a (IR subtraction) counterterm whose explicit expression \cite{Bozzi:2005wk}
is obtained from the resummation program of the logarithmically-enhanced
contributions to $q_T$ distributions. 
The `coefficient' ${\cal H}^{F}_{(N)NLO}$, which also compensates for the subtraction
of $d{\sigma}^{CT}_{(N)LO}$,
corresponds to the (N)NLO truncation of the process-dependent perturbative function
\begin{equation}
{\cal H}^{F}=1+\frac{\alpha_{\mathrm{S}}}{\pi}\,
{\cal H}^{F(1)}+\left(\frac{\alpha_{\mathrm{S}}}{\pi}\right)^2
{\cal H}^{F(2)}+ \dots \;\;.
\end{equation}
The NLO calculation  of $d{\sigma}^{F}$ 
requires the knowledge
of ${\cal H}^{F(1)}$, and the NNLO calculation also requires ${\cal H}^{F(2)}$. The general 
structure of ${\cal H}^{F(1)}$
is explicitly known;\cite{deFlorian:2000pr} exploiting the explicit results of ${\cal H}^{F(2)}$ for Higgs
\cite{Catani:2007vq,Catani:2011kr} and vector boson \cite{Catani:2009sm} 
production we have generalized the process-independent relation of Ref.\cite{deFlorian:2000pr} to 
the calculation of the NNLO coefficient 
${\cal H}^{F(2)}$.
%
%
%
\section{Quantitative results}
We have performed our fully-differential NNLO calculation~\cite{Catani:2011qz} of diphoton production
according to Eq.~(\ref{main}).
The NNLO computation is encoded
in a parton level
Monte Carlo program, in which
we can implement arbitrary IR safe cuts on the final-state
photons and the associated jet activity. 
We concentrate on the direct production of diphotons, and 
we rely on the smooth cone isolation criterion.\cite{Frixione:1998jh} Considering a cone of radius $r=\sqrt{(\Delta \eta)^2+(\Delta \phi)^2}$ around
each photon, we require the total amount of hadronic (partonic) transverse energy $E_T$ 
inside the cone to be smaller than $E_{T\, max}(r)$,
\begin{equation}
\label{eq:etmax}
E_{T}<E_{T\, max}(r) \equiv  \epsilon_\gamma \,p_T^\gamma \left(\frac{1-\cos r}{1- \cos R}\right)^{n}\,,
\end{equation}
where $p_T^\gamma$ is the photon transverse momentum; the isolation criterion
$E_T < E_{T\, max}(r)$ has to be fulfilled for all cones with $r\leq R$.
We use the MSTW 2008 \cite{Martin:2009iq} sets of parton distributions, with
densities and $\alpha_{\mathrm{S}}$ evaluated at each corresponding order,
and we consider $N_f=5$ massless quarks/antiquarks and gluons in 
the initial state. The default
renormalization ($\mu_R$) and factorization ($\mu_F$) scales are set to the value
of the invariant mass of the diphoton system,
$\mu_R=\mu_F = M_{\gamma\gamma}$. The QED coupling constant $\alpha$ is fixed to $\alpha=1/137$.\\
%
%
%
\begin{figure}[htb]
\begin{center}
\begin{tabular}{lr}
\psfig{figure=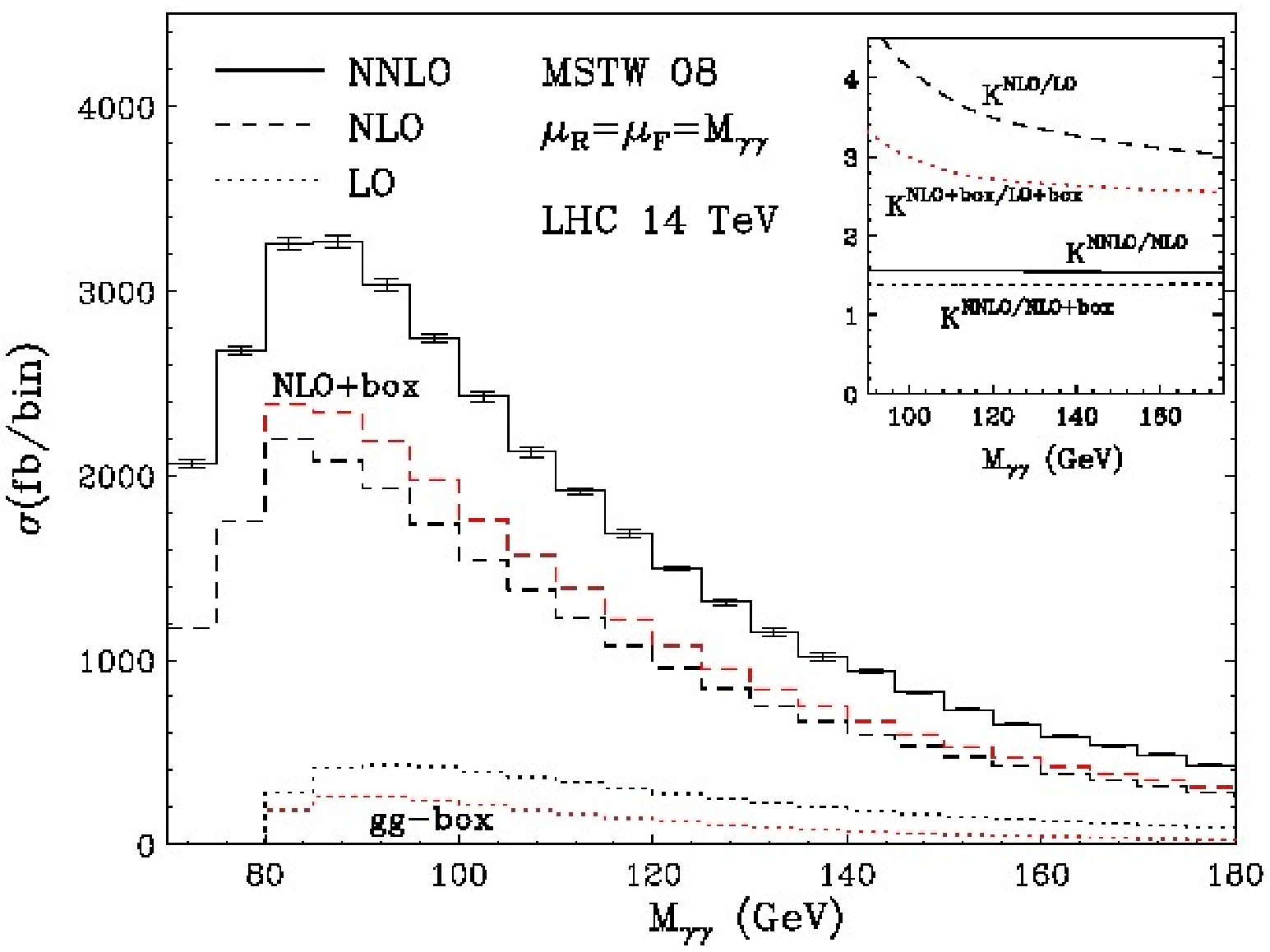,height=55mm,width=75mm}&
\psfig{figure=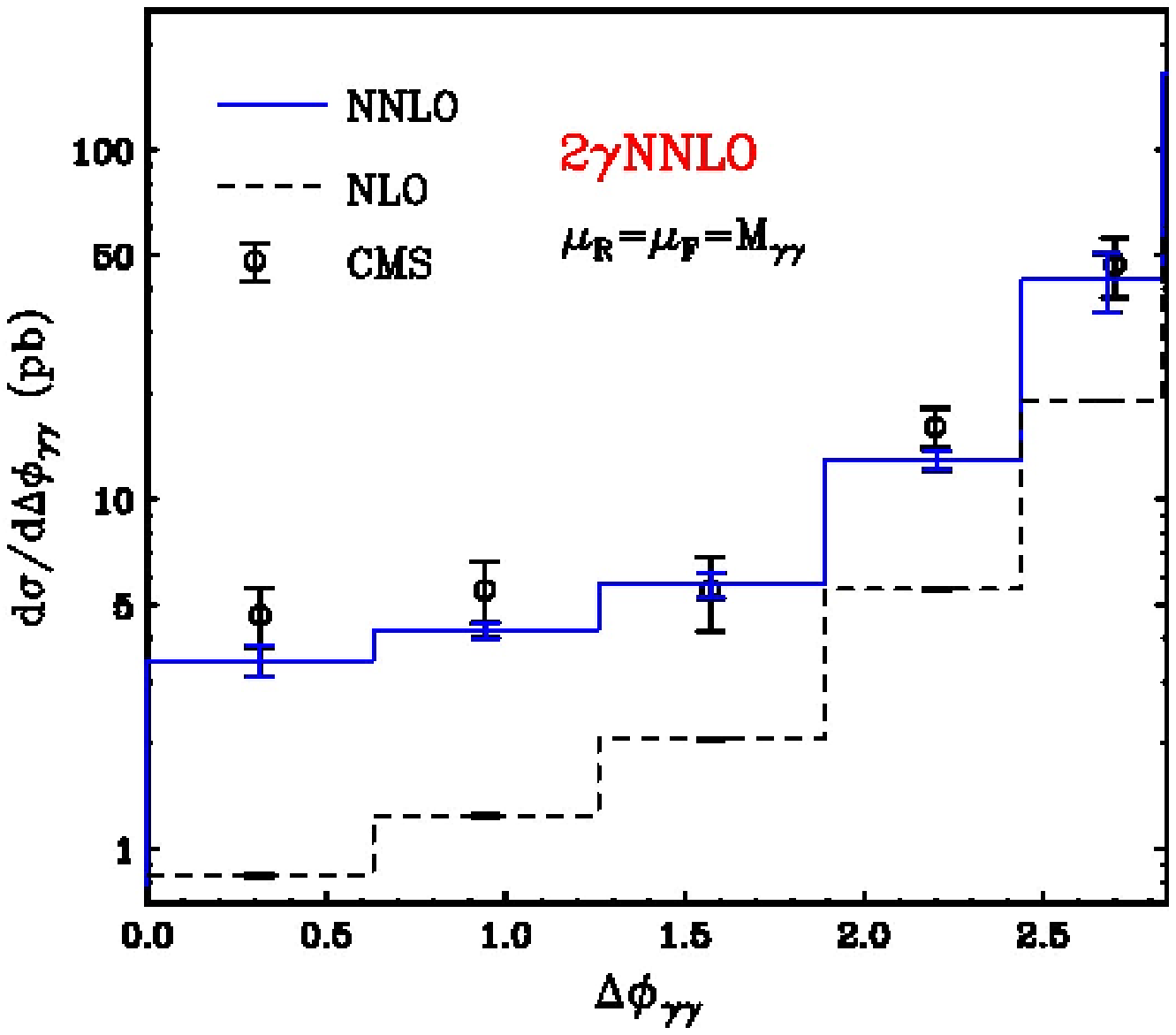,height=55mm,width=75mm}
\end{tabular}
\end{center}
\caption{\label{fig:mass}
{\em Left: Invariant mass distribution of the photon pair
at the LHC 
($\sqrt{s}=14$~{\rm TeV}): LO (dots), NLO (dashes) and NNLO (solid) results. We
also present the results of the box and NLO+box contributions. The inset plot
shows the 
corresponding {\rm K}-factors.~Right: Diphoton 
cross section as a function of the azimuthal separation of
  the two photons. Data from CMS~\protect\cite{Chatrchyan:2011qt} ($\sqrt{s}=7$~{\rm TeV}) are
  compared to the NNLO calculation.\protect\cite{Catani:2011qz}}}
\end{figure}
%
To present some quantitative results, we consider diphoton production
at the LHC ($\sqrt{s}=14$~TeV). We apply typical kinematical cuts used by ATLAS and CMS 
Collaborations in their Higgs search studies. We
require the harder and the softer photon to have transverse momenta $p_T^{\rm harder}\geq
40$~GeV and $p_T^{\rm softer}\geq 25$~GeV, respectively.
The rapidity of both photons is restricted to $|y_\gamma| \leq 2.5$, and
the invariant mass of the 
diphoton system is constrained to 
lie in the range $20 \,{\rm GeV}\leq M_{\gamma\gamma} \leq 250\,{\rm GeV}$.
The isolation parameters
are set to the values $\epsilon_\gamma=0.5$, $n=1$ and $R=0.4$. 
%
%
We observe \cite{Catani:2011qz} that 
the value of the cross section remarkably increases with the perturbative order
of the calculation. This increase is mostly due to the use of {\it very
asymmetric} (unbalanced) cuts on the photon transverse momenta. At the LO,
kinematics implies that the two photons are produced with equal transverse
momentum and, thus, both photons should have $p_T^{\gamma}\geq 40$~GeV. 
At higher orders, the final-state radiation of additional partons opens a new
region of the phase space, 
where $40$~GeV $\geq p_T^{\rm softer}\geq 25$~GeV. Since photons can copiously be
produced with small transverse momentum,\cite{Catani:2011qz} the cross section receives a sizeable
contribution from the enlarged phase space region. This effect is further
enhanced by the opening of a new large-luminosity partonic channel at each
subsequent perturbative 
order. 
%
%
In Fig.~\ref{fig:mass} (left) we compare the LO, NLO and NNLO invariant mass
distributions at the default scales. 
 The inset plot shows the K-factors defined as 
the ratio of the cross sections at two subsequent perturbative orders.
We note that ${\rm K}^{NNLO/NLO}$ is sensibly smaller than ${\rm K}^{NLO/LO}$,
and this fact indicates 
an improvement in the convergence of the perturbative expansion. 
We find that about 30\% of the NNLO corrections is due to the $gg$ channel (the 
{\it box contribution} is responsible for more than half of it), while almost
60\% still arises from the next-order corrections to the $qg$ channel.
\\
\\
Recent results from the LHC~\cite{Chatrchyan:2011qt,Aad:2011mh} and the
Tevatron \cite{Aaltonen:2011vk} show some discrepancies between the data
and NLO theoretical calculations of diphoton production. Basically, discrepancies were found in
kinematical regions where the NLO calculation is {\em effectively} a LO theoretical 
description of the process. Such 
phase space regions~\footnote{Away from the back-to-back configuration.} are accesible at 
NLO for the first time, due to the final-state radiation of the 
additional parton.\footnote{The low-mass
region ($M_{\gamma\gamma}\leq 80GeV$) in Figure \ref{fig:mass} also belongs to this case.} Figure~\ref{fig:mass} (right) 
shows a measurement by CMS,\cite{Chatrchyan:2011qt} of the diphoton
cross section as a function of the azimuthal angle $\Delta \phi_{\gamma\gamma}$ between the
photons. The data are compared with our NLO and NNLO calculations.\cite{Catani:2011qz}
The acceptance criteria used in this analysis ($\sqrt{\rm s}=7$~TeV) require of: $p_T^{\rm harder}\geq
23$~GeV and $p_T^{\rm softer}\geq 20$~GeV.
The rapidity of both photons is restricted to $|y_\gamma| \leq 2.5$, and
the invariant mass of the diphoton system is constrained to be $M_{\gamma\gamma} > 80\,{\rm GeV}$. The isolation parameters
are set to the values $\epsilon_\gamma=0.05$, $n=1$ and $R=0.4$. We note that the CMS data
are selected by using the standard cone isolation criterion and the constraint in Eq.~(\ref{eq:etmax}) is
applied only to the cone of radius $r=R$. Since the smooth isolation criterion used in our
calculation (we apply Eq.~(\ref{eq:etmax}) for all cones with $r\leq R$) is stronger than the photon
isolation used by CMS, we remark that our NLO and NNLO results cannot overestimate the corresponding
theoretical results for the CMS isolation criterion. The histograms in Fig.~\ref{fig:mass} (right) show
that the NNLO QCD results remarkably improve the theoretical description of the CMS data throughout the 
entire range of $\Delta \phi_{\gamma\gamma}$. 
%
%
\\
\\
The results illustrated in this contribution show that the NNLO description of diphoton 
production is essential to understand the 
phenomenology associated to this process, and therefore, the NNLO calculation is a relevant tool 
to describe the main background for Higgs boson searches.

\medskip

\textbf{Acknowledgments}

I would like to thank Stefano Catani and Daniel de Florian for helpful comments, 
and to Ed Berger for inviting me to give this talk. This work was 
supported by the INFN and the Research Executive Agency (REA) of the European Union under the Grant Agreement number PITN-GA-2010-264564 (LHCPhenoNet).


\end{document}